# Electrical reversal of the magnon thermal Hall coefficient in a van der Waals bilayer antiferromagnet

Yu-Hao Shen[1,2], Yiqun Liu[1,2]*, Min Luo[3]

[1]National Laboratory of Solid-State Microstructures and School of Physics, Nanjing University, Nanjing 210093, China

[2]Collaborative Innovation Center of Advanced Microstructures, Nanjing University, Nanjing 210093, China

[3]School of Information and Communication Engineering, Shanghai Technical Institute of Electronics & Information, Shanghai, 201411, China

*Author to whom any correspondence should be addressed.
E-mail: liuyiqun@mail.bnu.edu.cn

## ABSTRACT

With spin-layer locking, the manipulation of spin degree of freedom via perpendicular electric field can be realized in a typical antiferromagnetically coupled bilayer. In analogy to the electric control of the anomalous layer Hall effect of electron within such bilayer system, we propose here its magnon counterpart i.e., thermal Hall effect controlled by a perpendicular electric field. Unlike electrons, magnon is charged neutral and its transport in solids can be driven by a thermal gradient. It also exhibits Hall response due to the intrinsic Berry curvature of magnon, analogous to the achievement in electron system. Taking bilayer 2H-VSe$_2$ with both H-type stacking and interlayer antiferromagnetic coupling as a platform, we perform first-principles calculations towards the magnetic exchange coupling parameters under applied electric field perpendicular to the plane. Based on linear spin wave approximation, we then fit the magnon band structures accordingly and calculate the corresponding Berry curvature. The thermal Hall coefficient dependence on the temperature under thermal gradient can be calculated correspondingly in linear response regime. It is shown that electric field reversal is able to reverse the sign of the coefficient. These findings provide a platform for the realization of all-electric magnon spintronics.

**INTRODUCTION**

Two-dimensional (2D) van der Waals layered materials provide ideal platforms for the investigation towards the interplay among multiple degrees of freedom of the quasiparticle. Typical studies about electron system with good quantum numbers of charge, spin, valley as well as layer degree of freedom in such systems has recently become a great of interest[1-5]. Besides, the interlayer exciton in a bilayer heterojunction, that is, low energy particle-hole excitation within different individual layer, has also attracted intensive attention[6-9]. Importantly, the interlayer tunneling have considerable influence on the interlayer excitation. Also, by applying a perpendicular electric field, which breaks the potential balance between layers, brings about tunable physical properties associated with both intralayer and interlayer quasiparticle excitation. Among bilayer heterostructures, of particular interest is the antiferromagnetically coupled two ferromagnetic monolayer system. It possess spin-layer locking and can achieve anomalous layer Hall effect[10,11] and anomalous valley Hall effect[12,13] for electron controlled by perpendicular electric field. However, the physical motion of charges unavoidably give rise to Joule heating for such electronic system. It is natural to consider a charged neutral magnon excitation for the purpose of obviating dissipation from Joule heating.

As a quasiparticle of spin wave excitation in magnetically ordered system, magnon can transfer both energy and spin over centimeter distance within a magnetic insulator[14,15]. Interestingly, it is widely recognized that in a 2D collinear antiferromagnet exhibiting Néel order along the easy-axis of magnetic anisotropy[16-18], the magnon coherent states can be locked with opposite spin angular momentum characterized by opposite chirality. These can manifest as either left-handed or right-handed precession modes relative to the magnetization axis and be selectively excited and detected[19-22], enabling its coding with binary information similar to the electron spin-1/2 states. For a typical honeycomb lattice, which is staggered with two triangular sublattices possessing opposite perpendicular magnetization[18,23], the two chiral modes can undergo transverse transport under a longitudinal thermal gradient. This transport behavior is driven by their intrinsic Berry curvature, analog to the Hall response observed in electrons. As a consequence of Dzyaloshinskii-Moriya interaction (DMI)[23-25], there lifts their degeneracy and gains spin Hall current i.e., magnon spin Nernst effect, which has been observable experimentally[26]. It is noteworthy that there vanishes the heat Hall current owing to symmetry arguments. If the two sets of sublattices are divided into separate layers, as depicted in Fig. 1(a), a straightforward approach would be to apply an electric field perpendicular to the 2D plane. This would break the symmetry, resulting in distinct magnetic exchanges between the two layers. Thus, there will lead to nonvanishing heat transport of the system that can be electrically tunable for such bilayer system.

Along this strategy, in this paper we take bilayer 2H-VSe$_2$ with H-type stacking as a platform, and set interlayer antiferromagnetic coupling. The degree of freedom of sublattice A, B in a honeycomb monolayer is now equivalent to the layer degree of freedom a, b in such bilayer. And we perform first-principles calculations towards the magnetic exchange coupling parameters under applied electric field perpendicularly. Then we fit the magnon band structures from these calculated magnetic parameters in the context of linear spin wave approximation utilizing the spin Hamiltonian. It is found that there possesses splitting of the two chiral magnon bands, denoted as $\alpha$ and $\beta$, that comes from both internal DMI and the exchange coupling difference induced by the external electric field. It turns out that the magnon energy $\omega_\alpha(\pm q) \neq \omega_\beta(\mp q)$, causing a population imbalance between the $q$ and $-q$ states. Through calculating the magnon Berry curvature and hence the thermal Hall coefficient in linear response regime, we show that electric field reversal enables the sign change of this coefficient and there will reverse the thermal voltage between the two Hall bars along the sample's boundaries correspondingly.

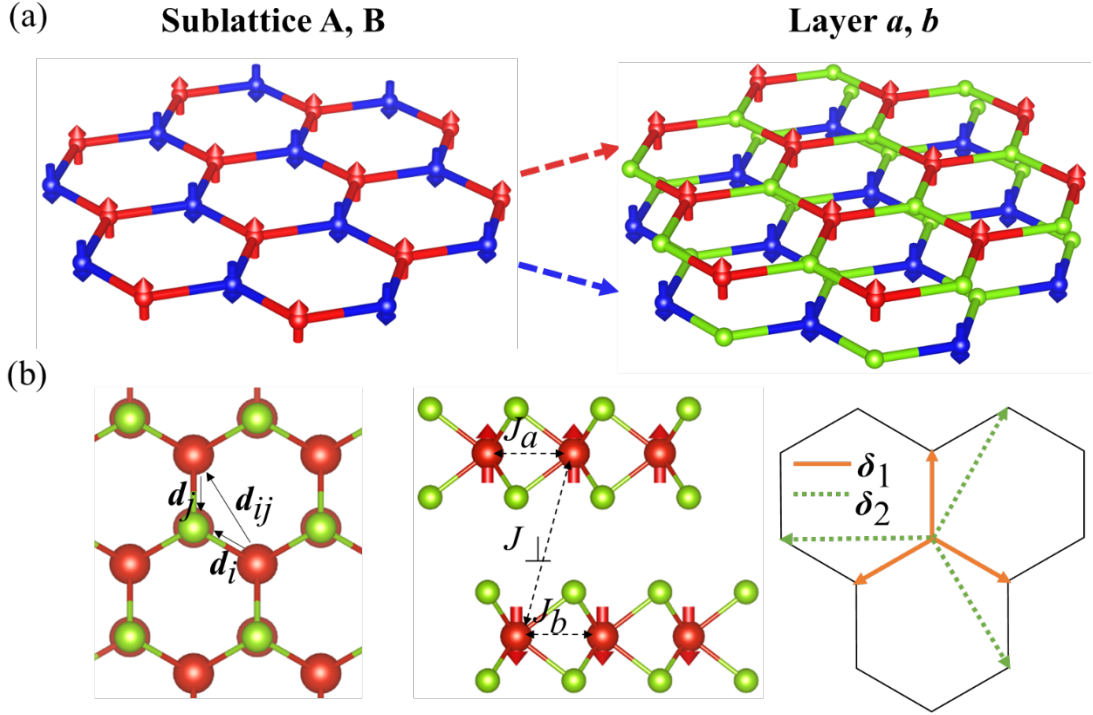

**Figure. 1** (a) Scheme of the generalization to the antiferromagnetically coupled bilayer system from an antiferromagnetic monolayer with honeycomb lattice, where the degree of freedom of sublattice A, B in a honeycomb monolayer is now equivalent to the layer degree of freedom a, b in such bilayer. (b) Top view (left panel) and side view (middle panel) of H-type stacked bilayer 2H-VSe$_2$. The magnetization axis is set to be perpendicular to the 2D plane. In the right panel, we show the lattice vector that connects nearest ($\delta_1$) and next-nearest ($\delta_2$) sites, denoted by dashed arrows, in a honeycomb lattice.

## COMPUTATIONAL METHOD AND THEORY

### A. First-principles calculations

For the H-type stacked bilayer 2H-VSe$_2$ with interlayer antiferromagnetic coupling, the optimized lattice constant is 3.335 Å and the interlayer distance between the two Se layers is 3.220 Å. The first-principles calculations are performed with density functional theory (DFT) using the projector augmented wave (PAW) method implemented in the Vienna ab initio Simulation Package (VASP)[27]. The exchange-correlation potential is treated in Perdew-Burke-Ernzerhof form[28] of the generalized gradient approximation (GGA-PBE) with a kinetic-energy cutoff of 400 eV. Further, we use GGA+U scheme and set the effective Hubbard U repulsion $U_{eff}$ = 1 eV to correct electron-electron correlation. A well-converged 11×11×1 Monkhorst-Pack $k$-point mesh is chosen in self-consistent calculations of the unit cell. The convergence criterion for the electronic energy is $10^{-5}$ eV and the structures are relaxed until the Hellmann-Feynman forces on each atom are less than 1 meV/Å.

The exchange and DMI parameters can be calculated by the four-state method[29,30] based on the spin Hamiltonian. In the supercell static calculations needed for four-states method, we use a 3×3×1 supercell and 5×5×1 $k$-point sampling. For our calculations, the spin-orbit coupling effect was included and the dispersion corrected DFT-D2 method[31] is adopted to describe the van der Waals interactions. The external electric field is introduced with the planar dipole layer method.

### B. Spin Hamiltonian of bilayer 2H-VSe$_2$ system

The individual monolayer of 2H-VSe$_2$ possess a typical honeycomb lattice of V atomic plane sandwiched between two Se atomic planes (Fig. 1(b)). As a kind of ferrovalley system for the monolayer, it exhibits spontaneous valley polarization and become an anomalous valley Hall (AVH) insulator[32-34]. For the antiferromagnetically coupled bilayer realized as an antiferrovalley system, the AVH can be electrically controlled via interlayer gate bias[13,35]. These achievements in practice lies on that the 2H-VSe$_2$ is predicted to be a room-temperature 2D ferromagnet with transition temperature $T_c$ ~ 590 K [36]. And for the bilayer case, it is found to possess interlayer antiferromagnetism[35,37]. Once the intralayer ferromagnetic order is formed, there exists large sizable domain in individual layer for the interlayer coupling to remain antiferromagnetic[35]. Moreover, VSe$_2$ monolayer system, as a 2D ferromagnet, usually displays an easy plane single-ion magnetic anisotropy demonstrated by many first-principles calculations and also evaluated experimentally[34,38,39]. Yet its magnetization perpendicular to the plane can also be stabilized since the magnetic anisotropy energy can be tuned externally[38,40,41]. Then below for a H-type stacked bilayer antiferromagnet system, which is energetically favorable compared with other types of interlayer stacking, we assume the system undergoes a small perpendicular magnetic anisotropy.

We start by mapping the magnetic energy from a minimal spin Hamiltonian for our bilayer antiferromagnet[42], which is similar to model antiferromagnetic honeycomb lattice[18,23,29,43-45]:

$$H = J_\perp \sum_{\langle ij \rangle} \mathbf{S}_{i,a} \cdot \mathbf{S}_{j,b} + \sum_l J_l \sum_{\langle ij \rangle} \mathbf{S}_{i,l} \cdot \mathbf{S}_{j,l} + \sum_l \sum_{\langle ij \rangle} \mathbf{D}_{ij,l} \cdot \mathbf{S}_{i,l} \times \mathbf{S}_{j,l} - K \sum_{i,l} (S_{i,l}^z)^2 \quad (1)$$

Here, $J_\perp > 0$ is the interlayer antiferromagnetic exchange between nearest neighbors and $J_l < 0$ is the intralayer ferromagnetic one. For the intralayer coupling, the *ij* connects the magnetic ions within triangular lattice. The DMI vector survives within the intralayer triangular lattice, denoted as $\mathbf{D}_{ij,l}$, and vanishes for the interlayer nearest neighbors because of the presence of the inversion center at the midpoint. Besides, different from the antiferromagnetic monolayer case, its planar components $D_{ij,l}^{x,y}$ can be nonzero in such bilayer. However, for such a magnetic ground state of Néel order along $z$ direction, within linear spin wave approximation we will use below, the linear terms of magnon creation or annihilation cancel and no quadratic terms exists[46]. For simplicity we only perform calculations towards the perpendicular components $D_{ij,l}^z$ and specifically it can be expressed as $D_{ij,l}^z = v_{ij} D_l^z$, where $v_{ij} = \mathbf{d}_i \times \mathbf{d}_j \cdot \hat{\mathbf{z}} / |\mathbf{d}_i \times \mathbf{d}_j|$ based on symmetry requirements. The lattice vector $\mathbf{d}_i$, $\mathbf{d}_j$ and $\mathbf{d}_{ij} = \mathbf{d}_i - \mathbf{d}_j$ are shown in Fig. 1(c) that connects nearest and next-nearest neighbors respectively in the honeycomb lattice of each individual layer. The last term refers to a perpendicular anisotropic term with $K > 0$ in both *a, b* layer.

### C. Magnon bands

Using the Holstein-Primakoff (HP) transformation[47] for lattice spin $\mathbf{S}_i$ with neglecting magnon-magnon interaction[46,48]:

$$S_{i,a}^{+} \simeq \sqrt{2S} a_i, \quad S_{i,a}^{-} \simeq \sqrt{2S} a_i^{\dagger}, \quad S_{i,a}^{z} = S - a_i^{\dagger} a_i$$

$$S_{i,b}^{+} \simeq \sqrt{2S} b_i^{\dagger}, \quad S_{i,b}^{-} \simeq \sqrt{2S} b_i, \quad S_{i,b}^{z} = b_i^{\dagger} b_i - S \tag{2}$$

Here $a_i^{+}(a_i)$ and $b_i^{+}(b_i)$ corresponds to the magnon bosonic creation (annihilation) operator on site $i$ in layer $a$ and $b$, respectively. After taking Fourier transformation:

$$a_q = \frac{1}{\sqrt{N}} \sum_i a_i e^{iq \cdot R_i}, \quad a_q^{\dagger} = \frac{1}{\sqrt{N}} \sum_i a_i^{\dagger} e^{-iq \cdot R_i}$$

$$b_q = \frac{1}{\sqrt{N}} \sum_i b_i e^{-iq \cdot R_i}, \quad b_q^{\dagger} = \frac{1}{\sqrt{N}} \sum_i b_i^{\dagger} e^{iq \cdot R_i} \tag{3}$$

there gives (disregarding constant terms):

$$H = S \sum_q (M_q a_q^{\dagger} a_q + N_q b_q^{\dagger} b_q + T_q^{*} a_q^{\dagger} b_q^{\dagger} + T_q a_q b_q) \tag{4}$$

where $M_q = 3J_{\perp} + K(2S-1)/S - 2J_a(3-\gamma_q) + 2D_a^z \lambda_q$, $N_q = 3J_{\perp} + K(2S-1)/S - 2J_b(3-\gamma_q) - 2D_b^z \lambda_q$ and $T_q = J_{\perp} t_q$ with $\gamma_q = \sum_{\delta_2} \cos(q \cdot \delta_2)$, $\lambda_q = \sum_{\delta_2} \sin(q \cdot \delta_2)$ and $t_q = \sum_{\delta_1} e^{iq \cdot \delta_1} = |t_q| e^{i\varphi_q}$. Following previous studies about such antiferromagnetic case[23], we use Bogoliubov quasiparticle representation $\alpha_q = u_q a_q - v_q b_q^{\dagger}$ and $\beta_q = u_q b_q - v_q a_q^{\dagger}$ to diagonalize $H_q$ (see Appendix A).

### D. Two magnon chiral modes

It is evident that the spin z-component $S^z$ at each site remains conserved in the magnetic ground state, specifically the colinear antiferromagnetic order aligned along the $z$-direction. This conservation is due to the rotational symmetry around the z-axis in spin space, denoted as $c_z$, which is preserved by the spin Hamiltonian. Consequently, the DMI term disappears, satisfying the condition $[S^z, H] = 0$. In consideration of spin wave excitation at low temperature, note that DMI appears, and for each site $S^z$ is not conserved. However, the total spin:

$$S^z = \sum_q (b_q^{\dagger} b_q - a_q^{\dagger} a_q) = \sum_q (\beta_q^{\dagger} \beta_q - \alpha_q^{\dagger} \alpha_q) \tag{5}$$

It can be directly proven that $[S^z, H] = 0$, indicating that within the linear spin wave approximation, the total spin angular momentum $S^z$ along the easy-axis remains a good quantum number for magnons, even under the influence of a perpendicular electric field. This electric field solely affects the magnetic parameters without breaking the rotational symmetry around the $z$-axis, denoted as $c_z$. Moreover, since $\langle \alpha | -S^z | \alpha \rangle = \langle \beta | S^z | \beta \rangle = 1$, which give rise to opposite spin angular momentum $\pm \hbar$ carried by the two modes. Physically, these two magnon modes is locked with opposite chirality that their coherent state can be left-handed or right-handed precession modes with respect to the easy-axis.

### E. Magnon Berry curvature and thermal Hall coefficient

Since the eigenstates $\psi_q$ are exactly the same for $\alpha$ and $\beta$ mode, they share the same corresponding Berry curvature of magnon, which can be calculated as[23]:

$$\Omega_q = i \langle \nabla_q \psi_q | \times \sigma_z | \nabla_q \psi_q \rangle \tag{6}$$

Here, $\sigma_z$ factor comes from the generalized normalization conditions $\langle \psi_q | \sigma_z | \psi_q \rangle = 1$. In the linear response regime[49-51], when a thermal gradient $\nabla T$ is uniformly applied along a planar direction of the 2D samples over a sufficiently large area, a transverse thermal current arises[52-54]. Here, the thermal Hall coefficient $\kappa_{xy}$ is directly proportional to the intrinsic Berry curvature of the magnon:

$$\kappa_{xy} = \frac{k_B^2 T}{\hbar} \sum_{n=\alpha,\beta} \int \frac{d^2 q}{(2\pi)^2} \Omega_{nq} c_2(\rho_{nq}) \tag{7}$$

Here, $k_B$ is Boltzmann constant and $c_2(\rho_{nq}) = (1 + \rho_{nq})\left(\ln \frac{1+\rho_{nq}}{\rho_{nq}}\right)^2 - (\ln \rho_{nq})^2 - 2\text{Li}_2(-\rho_{nq})$, where $\text{Li}_2(z)$ is the polylogarithm function and $\rho_{nq} = 1/(e^{\hbar \omega_n(q)/k_B T} - 1)$ is the Bose-Einstein distribution function with the chemical potential $\mu = 0$ due to that the magnon number does not conserve in consideration of inelastic scatterings[50]. The transverse thermal current of magnon is then expressed as $\boldsymbol{J}_\epsilon = \kappa_{xy} \hat{z} \times \nabla T$.

## RESULTS AND DISCUSSION

In the absent of perpendicular electric field applied, we extract the total energy and spin magnetic moment $m_s$ = 1.16 $\mu_B$ on each V ion from first-principles calculations and obtain those parameters as presented in Table. I.

The calculated results of the exchange parameters are in agreement with those in previous work[55]. Next, we apply an electric field $E_z$ perpendicular to the plane of the layer, penetrating from layer $b$ to layer $a$ defined as the positive direction. And during the evolution of $E_z$, the slight difference $\Delta m_s$ of the spin magnetic moment (about 0.01 $\mu_B$ when $E_z$ = 0.1 eV/Å) between layers and the variation of $m_s$ is ignored at present. The calculated magnetic parameters under $E_z$ are shown in Fig. 2. It is found that within the accuracy we fix, there exhibits much more significant change of $J_l$ in comparison with that of $J_\perp$ and $D_l^z$. Specifically, the increase and decrease induced by electric field of $J_a$ and $J_b$ respectively is about 50 times larger than those variations of $J_\perp$ and $D_l^z$ (these two parameters and their variations are almost at the same order of magnitude). A qualitative explanation for these observations comes from that there induces intralayer electronic polarization, as a dielectric response to a small perpendicular external field $E_z$ in such bilayer system[35,56]. Since the two magnetic layers are alternately stacked in an H-type configuration, the application of $E_z$ shifts the charge center of the bilayer, which is originally located at the inversion center when $E_z = 0$. This shift effectively creates an electric dipole. Consequently, this induces both perpendicular and planar charge redistributions within each individual layer. Notably, the planar redistribution significantly affects the overlap of wavefunctions for the magnetic ions within the triangular lattice, thereby influencing the exchange integral $J_l$.

|  | $J_\perp$ | $J_a$ | $J_b$ | $D_a^z$ | $D_b^z$ |
|---|---|---|---|---|---|
| Our results | 0.27 | -14.90 | -14.90 | 0.17 | 0.17 |
| DFT data in Ref.[55] | 0.25 | -11.53 | -11.53 | | |

**Table. I** Values of the magnetic parameters in the spin Hamiltonian, in units of meV.

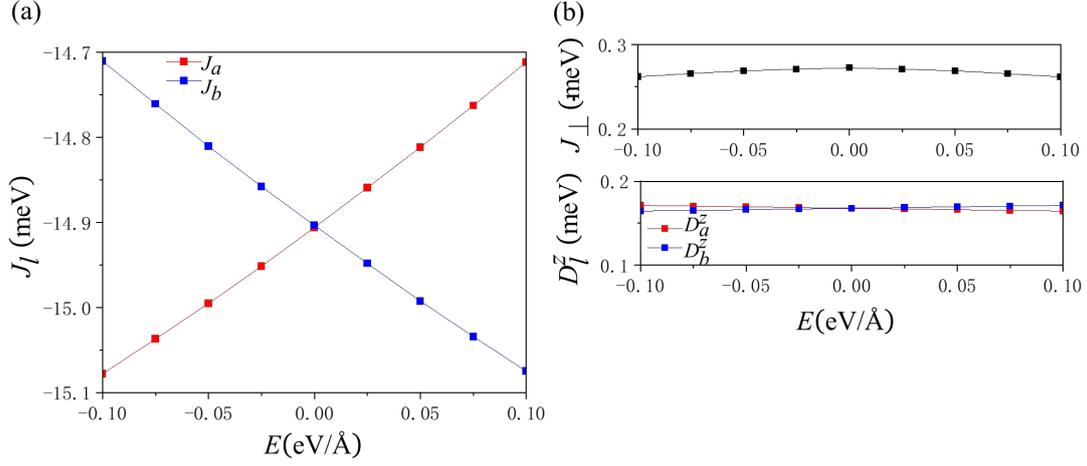

**Figure. 2** Calculated (a) planar exchange $J_l$ (b) perpendicular exchange $J_\perp$ and DMI parameter $D_l^z$ as the electric field $E$ evolves from -0.1 to 0.1, in units of eV/Å.

### A. Calculated magnon spectrum

We show the magnon band structure in Fig. 3 calculated from the magnetic exchange parameters above with setting $S=1$ and $K=1$ μeV. When $E_z=0$, as shown in Fig. 3(b), similar to the case of the antiferromagnetic monolayer[23,29], there satisfies $\omega_\alpha(\pm q) = \omega_\beta(\mp q)$ and here the bands splitting merely depends on the internal DMI. It can be shown in the case of in Fig. 3(a) and Fig. 3(c), where $E_z = \pm 0.025$ eV/Å, there leads to that $\omega_\alpha(\pm q) \neq \omega_\beta(\mp q)$. The leading contribution that results in this phenomenon comes from the exchange coupling difference $J_a - J_b$ in $M_q - N_q$ term of the magnon energy given by Eq. (A3).

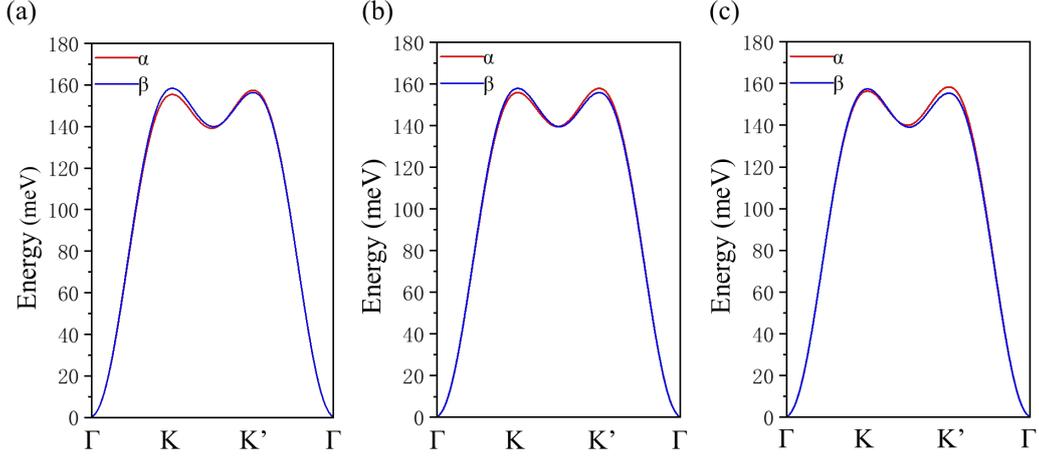

**Figure. 3** Magnon band structures along the high-symmetry line Γ-K-K'-Γ in the first Brillouin zone (BZ). They are calculated from the magnetic exchange parameters for (a) $E_z = -0.025$ eV/Å, (b) $E_z = 0$, (c) $E_z = 0.025$ eV/Å case.

## B. Symmetry arguments about the magnon spectrum

To give a symmetry argument, we then expand spin Hamiltonian in Eq. (1) to the quadratic order in $\mathbf{S}_a = \delta\mathbf{S}_a + S\hat{z}$ and $\mathbf{S}_b = \delta\mathbf{S}_b - S\hat{z}$ (setting $S = 1$ and disregarding constant terms)[23]:

$$H = J_\perp \sum_{\langle ab \rangle}(\delta\mathbf{S}_a \cdot \delta\mathbf{S}_b - \delta S_a^z + \delta S_b^z) + J_a \sum_{\langle aa' \rangle}(\delta\mathbf{S}_a \cdot \delta\mathbf{S}_{a'} - \delta S_a^z + \delta S_{a'}^z)$$
$$J_b \sum_{\langle bb' \rangle}(\delta\mathbf{S}_b \cdot \delta\mathbf{S}_{b'} - \delta S_b^z + \delta S_{b'}^z) + D_a^z \sum_{\langle aa' \rangle}(\delta S_a^x \delta S_{a'}^y - \delta S_a^y \delta S_{a'}^x)$$
$$-D_b^z \sum_{\langle bb' \rangle}(\delta S_b^x \delta S_{b'}^y - \delta S_b^y \delta S_{b'}^x) + K\sum_a[2\delta S_a^z + (\delta S_a^z)^2] + K\sum_b[-2\delta S_b^z + (\delta S_b^z)^2] \quad (8)$$

where, $\langle ab \rangle$ and $\langle aa' \rangle$ denotes the interlayer and intralayer nearest neighborhood respectively. It is easy to check that in the absent of both external electric field $E_z$ (then $J_a = J_b$ and $D_a^z = D_b^z$) and internal DMI,

$$\mathcal{T}c_x H (\mathcal{T}c_x)^{-1} = H \quad (9)$$

where the symmetry operations $\mathcal{T}c_x$ of combined time-reversal ($\mathcal{T}$) and a two-fold rotation around x axis in the spin space ($c_x$) act only on the spin deviation $\delta\mathbf{S}_a$ and $\delta\mathbf{S}_b$ while leaving the Néel ground state invariant. Based on HP transformation given in Eq. (2) and Fourier transformation given in Eq. (3), for modes $\alpha_q = u_q a_q - v_q b_q^\dagger$ and $\beta_q = u_q b_q - v_q a_q^\dagger$, we have

$$\mathcal{T}c_x \alpha_q (\mathcal{T}c_x)^{-1} = -\alpha_{-q}, \quad \mathcal{T}c_x \beta_q (\mathcal{T}c_x)^{-1} = -\beta_{-q} \tag{10}$$

where we use[57] $\mathcal{T}i\mathcal{T}^{-1} = -i$ and $\mathcal{T}S_{a,b}^\pm \mathcal{T}^{-1} = -S_{a,b}^\mp$. Here, $S_{a,b}^\pm = \delta S_{a,b}^x \pm i\delta S_{a,b}^y$. Thus, under symmetry operations $\mathcal{T}c_x$ onto both left and right side of EOM $\hbar\omega_\alpha \alpha_q = [\alpha_q, H]$ and $\hbar\omega_\beta \beta_q = [\beta_q, H]$, it requires that $\omega_\alpha(q) = \omega_\alpha(-q)$ and $\omega_\beta(q) = \omega_\beta(-q)$. With DMI considered, owing to the DMI part in Eq. (8) is odd under $\mathcal{T}c_x$ hence this symmetry is broken so that $\omega(q) \neq \omega(-q)$, as can be seen from the bands splitting shown in Fig. 3(b).

Similar analysis can be used to the combined symmetry operations $\mathcal{P}c_x$, where $\mathcal{P}$ represents the inversion operation, enabling that $\mathcal{P}\delta S_a \mathcal{P}^{-1} = \delta S_b$. Note that Néel ground state is also invariant under $\mathcal{P}c_x$ operations. We find $H$ is $\mathcal{P}c_x$ invariant if external electric field $E$ switched off. And it can be easy to show that

$$\mathcal{P}c_x \alpha_q (\mathcal{P}c_x)^{-1} = \beta_{-q}, \quad \mathcal{P}c_x \beta_q (\mathcal{P}c_x)^{-1} = \alpha_{-q} \tag{11}$$

Therefore, as expected, $\omega_\alpha(\pm q) = \omega_\beta(\mp q)$ for the case of $J_a = J_b$ and $D_a^z = D_b^z$ (Fig. 3(b)). As $E_z$ switched on, there induces $J_a \neq J_b$ and $D_a^z \neq D_b^z$. Hence, the $\mathcal{P}c_x$ symmetry is not present since it just exchanges the $\delta S_a$ and $\delta S_b$ part, not their corresponding magnetic parameters under such symmetry operations. Yet, the $\mathcal{T}c_x$ symmetry can be preserved as long as DMI is not included. In this manner, we argue that $\omega_\alpha(\pm q) \neq \omega_\beta(\mp q)$ in the absent of $\mathcal{P}c_x$ symmetry for the case of $E_z \neq 0$, shown in Fig. 3(a) and 3(c). Moreover, if we exchange both the magnetic parameters $J_a$ with $J_b$ and $D_a^z$ with $D_b^z$ under $\mathcal{P}c_x$ operations, H is invariant, which is indicated by the symmetry between the magnon band structures under $E_z$ and $-E_z$.

We also plot the calculated magnon Berry curvature in Fig. 4 for the case of $E_z = 0$. Notably, its antisymmetric distribution under $q \to -q$ relies on an effective time reversal symmetry for magnon. Under such symmetry operations, since $D_a^z = D_b^z$, $\theta_q = \theta_{-q}$ and $\varphi_{-q} = -\varphi_q$ so that $\Omega_{-q} = -\Omega_q$. It also reflects this symmetry is preserved for $\psi_q$ such that $\psi_{-q} = \psi_q^*$. When $E_z \neq 0$ hence $D_a^z \neq D_b^z$, because of the DMI term $D_a^z - D_b^z$ in $M_q + N_q$ introduced into the expression of Berry curvature, that anti-symmetry no longer exists. It is indicated by the effective time reversal symmetry is broken that $\psi_{-q} \neq \psi_q^*$.

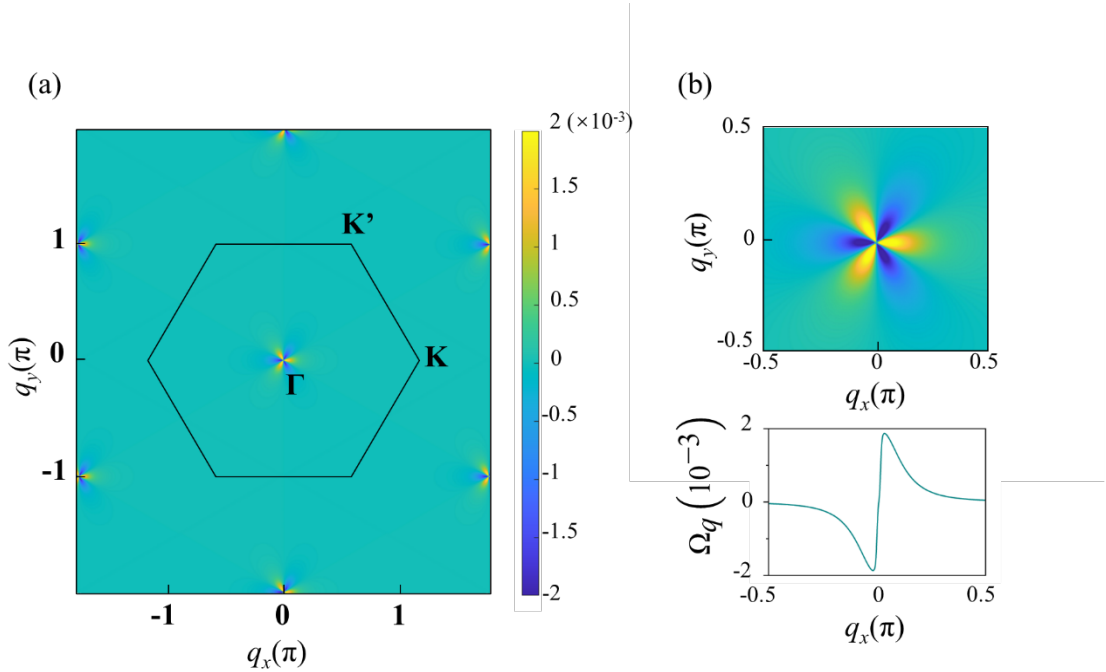

**Figure. 4** (a) Calculated Berry curvature $\Omega_q$ in the absent of perpendicular electric filed. The black hexagon denotes the 2D BZ boundary in wave vector $q$ space. (b) The distribution of $\Omega_q$ around Γ (upper panel) and along $q_x$ axis (lower panel).

### C. Electrical reversal of the magnon thermal Hall voltage

Given that $\omega_\alpha(\pm q) \neq \omega_\beta(\mp q)$ under perpendicular electric field $E_z$ switched on,

it is apparent that driven by a longitude thermal gradient, the thermal Hall transport cannot vanish identically caused by the distribution imbalance $\rho_q \neq \rho_{-q}$ of the thermal magnon, distinct from that case of $E_z = 0$. Clearly, earlier symmetry arguments show that when $E_z = 0$, $\Omega_{-q} = -\Omega_q$ and $\rho_\alpha(\pm q) = \rho_\beta(\mp q)$ so that $\kappa_{xy}$ vanishes due to the zero integral over BZ for an odd function. As plotted in Fig. 5(a), we reveal that the calculated thermal Hall coefficient $\kappa_{xy}$ from Eq. (7) can be nonzero if the perpendicular electric filed $E_z$ is switched on. Moreover, there possess the symmetry requirement that $\kappa_{xy}$ should be an odd function of the external field $E_z$, suggesting an electrical reversal of the coefficient $\kappa_{xy}$. In this sense, we achieve an all-electric control of thermal magnon transport in such bilayer antiferromagnet. To understand the temperature dependence, one can consider the contribution to the thermal Hall coefficient $\kappa_{xy}$ around Γ in BZ, which is primarily influenced by the distribution of the magnon Berry curvature. In this region, thermal magnons of long wavelengths predominate. The significant increase observed after approximately $T >$ 2.2 K can be attributed to adequate thermal excitation of magnons with a substantial Berry curvature. It is worth noting that as $q \rightarrow 0$, the Berry curvature rapidly diminishes for thermal magnons, as can be seen in Fig. 4(b). Moreover, the sign changes for specific finite $E$ case at about $T$ = 2.2 K, clearly shown in the inset figure of Fig. 5(a). This mainly because below 2.2 K, the calculated $\kappa_{xy}$ is dominated by the difference of Berry curvature between $q$ and $-q$ states, as we discuss for $E \neq 0$ case. Since there is almost no band splitting between $\alpha$ and $\beta$ modes below $k_BT \approx 0.2$ meV, the tiny Berry curvature difference of the thermal magnon contribute a tiny value of $\kappa_{xy}$. Above 2.2 K, the band splitting for thermal magnon modes dominates. And these two dominated contributions towards $\kappa_{xy}$ are opposite.

The thermal Hall voltage can be electrically reversed, as shown schematically in Fig. 5(b). Indeed, it is the magnon equivalent of the AVH effect observed in electronic system controlled by a perpendicular electric field $E_z$ [13]. In this scenario, the external electric field selects one of the carriers at the band edge, distinguished by its valley chirality, to attain anomalous velocities. Under $E_z$ and $-E_z$, there give rise to opposite transverse transport because of opposite Berry curvatures and hence the opposite valley charge accumulations on sample boundaries. In such a way, the AVH voltage can be electrically reversed. In analogy, the electric field $E_z$ in our magnon case can choose the net transverse transport between the two chiral magnon modes. The opposite Berry curvatures of the $q$ and $-q$ states give rise to contrasting energy accumulations in the transverse direction under $E_z$ and $-E_z$, respectively. Consequently, this leads to a reversal in the thermal Hall voltage, which can be practically measured. Thus, such antiferromagnetically coupled bilayer 2H-VSe$_2$ is also a good candidate to realize such magnon thermal Hall effect reversed by electric means.

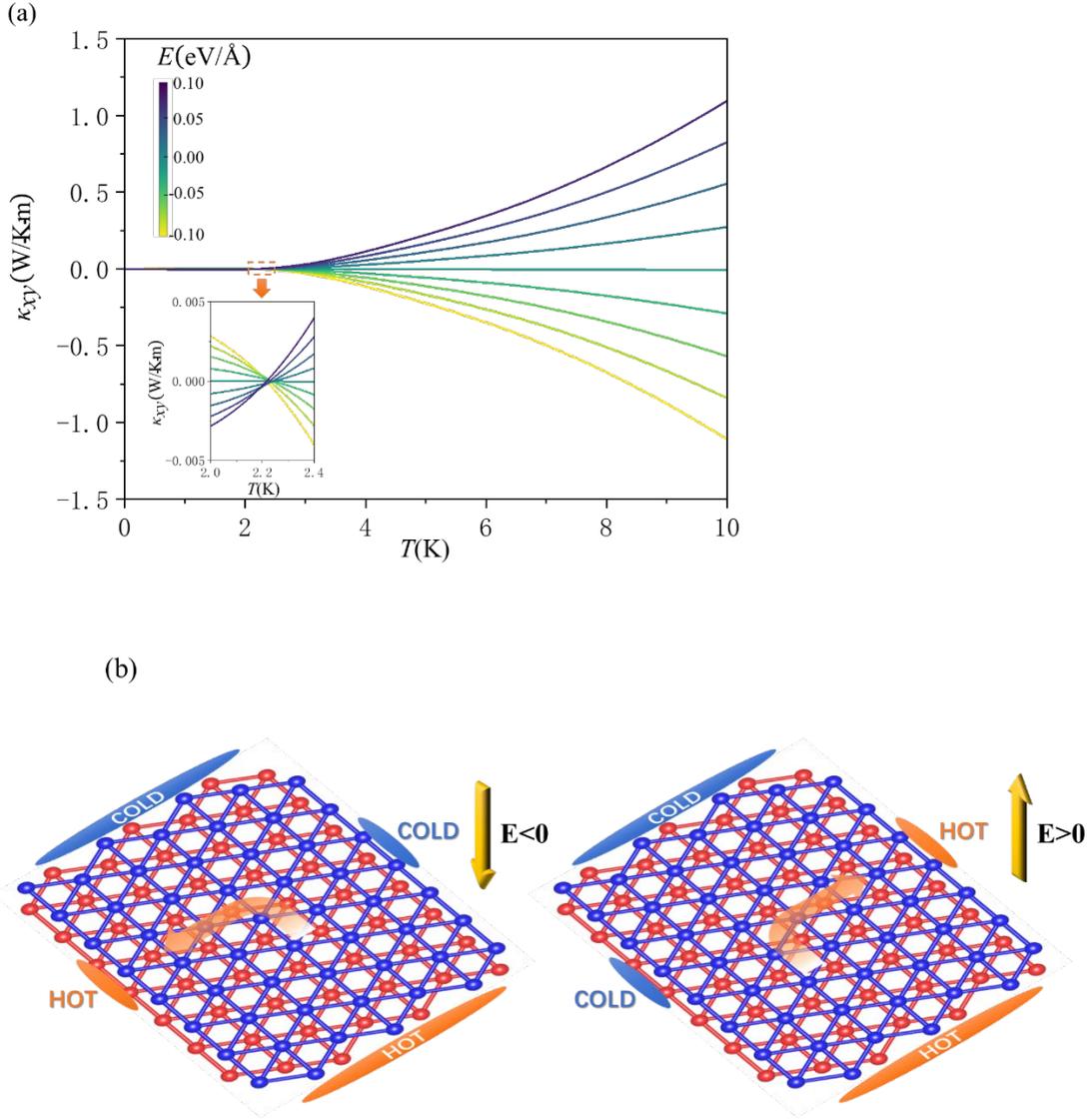

**Figure. 5** (a) The thermal Hall coefficient $\kappa_{xy}$ in units of W/Km versus temperature $T$. Here we use the distance between two magnetic layers $d$ = 6.35 Å for stacks of two-dimensional layers and divide by the interlayer distance $d$ to get the results (in units of W/Km). The inset figure shows the sign change of $\kappa_{xy}$ across about 2.2 K. (b) The electric reversal of the thermal Hall voltage. Here, the heater is applied along the longitude direction and we can measure the transverse temperature difference which indicates the sign of the Hall voltage.

## CONCLUSION

In summary, we theoretically investigate the magnon band structure for an antiferromagnetically coupled bilayer system with H-type stacking, which possesses a triangular magnetic lattice in each individual layer, in analogy to a honeycomb monolayer with Neel antiferromagnetic order. By breaking the symmetry between these two sublattice within the two layers through external perpendicular electric field we assumed, it is found that the magnon energy $\omega(q) \neq \omega(-q)$ mainly because of the exchange coupling difference of the two layers and hence the imbalance population for the thermal magnon between $q$ and $-q$ states. Subsequently, with nonzero Berry curvature calculated from the magnon eigen states we show the thermal Hall transport should occur under a longitude thermal gradient for such bilayer. Take 2H-VSe$_2$ bilayer antiferromagnet as a platform, we perform first-principles calculations towards the magnetic exchange coupling parameters under applied electric field perpendicularly. Then we fit the corresponding magnon band structures within linear spin wave approximation and calculate the Berry curvature. Through calculating the thermal Hall coefficient, it is shown that that we can reverse the sign of the thermal Hall voltage by the reversal of the external electric field, which is consistent with our symmetry arguments. Our findings open an appealing route toward functional thin film materials design for low energy consumption devices in all-electric magnon spintronics.

## ACKNOWLEDGMENTS

The work was supported by the Natural Science Foundation of Shanghai (Grant No. 19ZR1419800). We thank the National Supercomputer Center in Shenzhen.

## APPENDIX A: DIAGONALIZATION OF THE MAGNON HAMILTONIAN

Based on the equation of motion (EOM) $\hbar\omega_\alpha \alpha_q = [\alpha_q, H]$ and $\hbar\omega_\beta \beta_q = [\beta_q, H]$, we will get its matrix representation in the Nambu basis $\begin{pmatrix} a_q \\ b_q^\dagger \end{pmatrix}$ and $\begin{pmatrix} b_q \\ a_q^\dagger \end{pmatrix}$. Through

Pauli matrices $\sigma$ expansion form in particle-hole space, EOM can be written as:

$$\left(\cosh\theta_q \sigma_0 + \sinh\theta_q \cos\varphi_q \sigma_x + \sinh\theta_q \sin\varphi_q \sigma_y\right)\psi_q = \varepsilon_q \sigma_z \psi_q \tag{A1}$$

with $\psi_q = \begin{pmatrix} u_q \\ v_q \end{pmatrix}$ and hyperbolic parametrization:

$$\theta_q = \operatorname{arctanh}\frac{2|T_q|}{M_q + N_q}, \quad \varphi_q = \arg T_q, \quad \varepsilon_q = \frac{\frac{2\hbar\omega}{S} \pm (M_q - N_q)}{\sqrt{(M_q + N_q)^2 - 4|T_q|}} \tag{A2}$$

Here, '+' corresponds to the case of $\alpha$ mode and '−' corresponds to that of $\beta$ mode respectively. To solve the eigen problem for such a two-band system, we note that the EOM given by Eq. (A1) has the form $H\psi_q = \sigma_z \varepsilon_q$. And the diagonalization of $H$ gives identity matrix that $\langle \psi_q | H | \psi_q \rangle = 1$. The eigen value $\varepsilon_q = \pm 1$ corresponds to the positive and negative branch of the band structure with eigenstates that satisfy $\langle \psi_q | \sigma_z | \psi_q \rangle = \pm 1$ respectively. These two modes correspond to quasi-particle and quasi-hole excitations based on the bosonic commutation relations $[\alpha_q, \alpha_{q'}^+] = \pm \delta_{qq'}$. Now leaving out the negative one, we have the magnon eigen energy and eigen states as:

$$\hbar\omega_{\alpha,\beta}(\boldsymbol{q}) = S\left(\sqrt{\left(\frac{M_q + N_q}{2}\right)^2 - |T_q|^2} \mp \frac{M_q - N_q}{2}\right), \quad \psi_q = \begin{pmatrix} \cosh\frac{\theta_q}{2} \\ -\sinh\frac{\theta_q}{2} e^{i\varphi_q} \end{pmatrix} \tag{A3}$$

Here $\psi_q$ are the same for $\alpha$ and $\beta$ mode.